\documentclass[preprint,showpacs,preprintnumbers,amsmath,amssymb,aps]{revtex4}

\usepackage{graphicx}% Include figure files
\usepackage{dcolumn}% Align table columns on decimal point
\usepackage{bm}% bold math
\newcommand{\vecvar}[1]{\mbox{\boldmath$#1$}}

\begin{document}

\preprint{PRESAT-8301}

\title{First-Principles Study on Leakage Current through Si/SiO$_2$ Interface}

\author{Tomoya Ono}
\email{ono@prec.eng.osaka-u.ac.jp}
\affiliation{Department of Precision Science and Technology, Osaka University, Suita, Osaka 565-0871, Japan}

\date{\today}% It is always \today, today,

\begin{abstract}
The relationship between the presence of defects at the stacking structure of the Si/SiO$_2$ interface and leakage current is theoretically studied by first-principles calculation. I found that the leakage current through the interface with dangling bonds is 530 times larger than that without any defects, which is expected to lead to dielectric breakdown. The direction of the dangling bonds is closely related to the performance of the oxide as an insulator. In addition, it is proved that the termination of the dangling bonds by hydrogen atoms is effective for reducing the leakage current.
\end{abstract}

\maketitle
\section{Introduction}
The relevant length scale in silicon-based electronic devices has been continuously decreasing and is currently few nanometers. The understanding of the structural properties of Si/SiO$_2$ interfaces on the atomic scale acquires an enhanced technological importance because the defects at Si/SiO$_2$ interfaces cause serious problems, such as the increased gate leakage current, the reduced threshold for dielectric breakdown, and oxide charging \cite{sze,cv}. These problems are known to degrade the dielectric reliability and device performance. SiO$_2$ films on a Si(001) substrate are known to be mostly amorphous. However, intensive and sustained scientific activities by various methods up to now have shown that a crystalline phase of the SiO$_2$ can be observed down to $\sim$ 10 \AA \hspace{2mm} from the Si/SiO$_2$ interface in some cases \cite{intefaceexp1,intefaceexp2,intefaceexp3,intefaceexp4}. Although the dangling bonds in the amorphous structure of the SiO$_2$ films exhibit the complex and certain effect on the leakage, the influence of the defect at the interface becomes more important because atomic-layer SiO$_2$ is necessary even under high-$k$ dielectrics.

The ideal stacking structure of Si/SiO$_2$ has an interface consisting of only silicon and oxygen atoms without any defects; however this structure is extremely difficult to be formed experimentally in the entire interface region. Therefore, the termination of Si-dangling bonds associated with O vacancies by some proper atoms is a very important technology. Hydrogen has been known to play a crucial role in the fabrication of high-quality Si/SiO$_2$ interfaces. However, the degradation of devices has been attributed to hydrogen releasing from dangling bonds by hot electrons and diffusing to the interface \cite{hydrelease1,hydrelease2}. When the SiO$_2$ layer becomes very thin ($\sim$ 15 \AA), these dangling bonds are expected to increase markedly the leakage current through thin oxides as well as to degrade the dielectric reliability due to the direct tunneling of electrons from the substrate to the gate electrode. Therefore, the relationship between the presence of the defects at the interfaces and leakage current is definitely of interest.

In this paper, a first-principles study on the leakage current of ultrathin gate oxides with a SiO$_2$ thickness of $\sim$ 14 \AA \hspace{2mm} is carried out. In many cases, the leakage current thorough the Si/SiO$_2$ interfaces has been discussed by the one-dimensional Wentzel-Kramers-Brillouin approximation and/or the method using empirical parameters, which do not have enough accuracy to examine the effect of the dangling bonds from an atomistic point of view. There are also a considerable number of the first-principles studies exploring the dielectric breakdown of SiO$_2$ films \cite{kang}, however the conclusive picture of the relationship between the leakage current and the presence of Si-dangling bonds at Si/SiO$_2$ interfaces is still missing. I found that the resultant leakage current of the interface with the dangling bonds is 530 times larger than that without the dangling bonds and the length of the charge distribution of the defect state in the direction perpendicular to the interface is related to the leakage current. In addition, the termination of Si-dangling bonds by hydrogen atoms is effective for reducing the leakage current.

The rest of this paper is organized as follows. In Sec. II, I briefly describe the computational methods and models used in this study. My results are presented and discussed in Sec. III. I summarize my findings in Sec. IV.

\section{Computational methods and models}
Our first-principles calculation method for obtaining the electron-conduction properties is based on the real-space finite-difference approach \cite{rsfd,icp,tsdg}, which enables us to determine the self-consistent electronic ground state with a high degree of accuracy by a timesaving double-grid technique \cite{icp,tsdg} and the direct minimization of the energy functional \cite{dm}. Moreover, the real-space calculations eliminate the serious drawbacks of the conventional plane-wave approach, such as its inability to describe nonperiodic systems accurately. I examine the electron-conduction properties of the Si/SiO$_2$ interface suspended between semi-infinite electrodes by computing the scattering wave functions continuing from one electrode to the other.

\begin{figure}[htb]
\begin{center}
\includegraphics{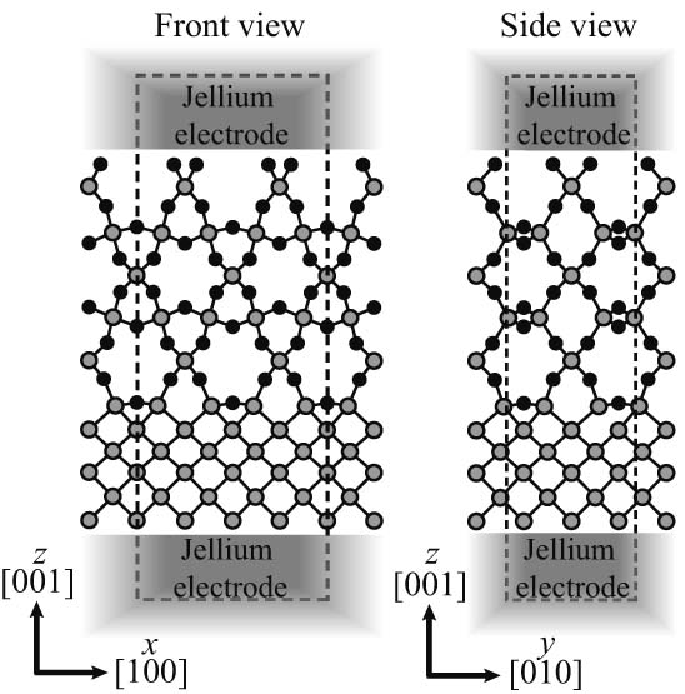}
\end{center}
\caption{Schematic image of computational model. Gray and black circles represent silicon and oxygen atoms, respectively. The rectangle enclosed by broken lines represents the scattering region when the leakage current is calculated.}
\label{fig:fig1}
\end{figure}

\begin{figure*}[htb]
\begin{center}
\includegraphics{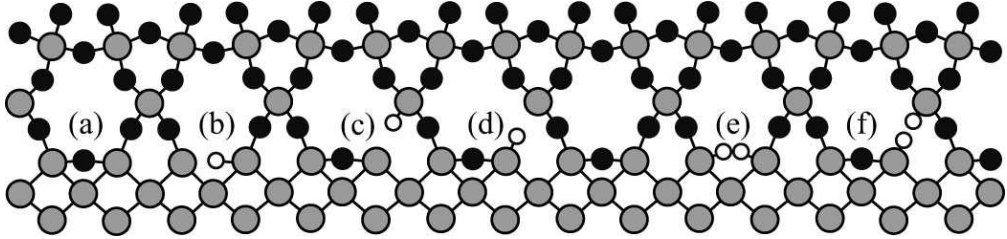}
\end{center}
\caption{Si/SiO$_2$ interface models. Perfect interface without any defects (a), hydrogen-atom bridge parallel to interface (b), $P_b$ center (c), hydrogen-atom bridge perpendicular to interface (d), hydrogen-molecule bridge parallel to interface (e), and hydrogen-molecule bridge perpendicular to interface (f). The meanings of the symbols are the same as those in Fig.~\ref{fig:fig1} and white circles represent hydrogen atoms.}
\label{fig:fig2}
\end{figure*}

Figure~\ref{fig:fig1} shows an example of the computational model for electron-conduction calculation. The Si/SiO$_2$ interface structure proposed by Kageshima and Shiraishi \cite{model} in which an $\alpha$-quartz crystalline SiO$_2$ layer stacked on a Si(001) surface is employed. From the first-principles analysis for the interface energy, the interface structure employed here is one of the candidates among the possible interface structures \cite{buczko}. The model shown in Fig.~\ref{fig:fig1} corresponds to an interface without any dangling bonds, which I refer to as the perfect interface hereafter. The thickness of the SiO$_2$ layers is $\sim$ 14 \AA \hspace{1mm} and that of the silicon substrate is 7.7 \AA. According to the oxidation process proposed by Kageshima and Shiraishi \cite{model}, the perfect interface appears after the emission of a silicon atom when six oxygen atoms are inserted into a silicon bulk. However, even when the number of inserted oxygen atoms is five, the silicon atom can be emitted and two dangling bonds remain. Since hydrogen is known to passivate the dangling bonds, I employ the interface models assuming that one oxygen atom at the interface is absent and the dangling bonds are terminated by hydrogen atoms. Figure~\ref{fig:fig2} shows the computational models: the perfect interface without any defects [model (a)], the hydrogen-atom bridge parallel to the interface [model (b)], $P_b$ center [model (c)], the hydrogen-atom bridge perpendicular to the interface [model (d)], the hydrogen-molecule bridge parallel to the interface [model (e)], and the hydrogen-molecule bridge perpendicular to the interface [model (f)]. Model (b) [models (c) and (d)] is expected to emerge after one hydrogen atom in model (e) [model (f)] is released from dangling bonds by hot electrons. The side length of the supercell is taken to be 10.86 \AA$\times$5.43 \AA \hspace{2mm} and {\it k}-space integrations are performed with $1\times2\vecvar{k}$ points in the irreducible wedge of a two-dimensional Brillouin zone. The norm-conserving pseudopotentials \cite{norm} of Troullier and Martins \cite{tmpp} are employed to describe the electron-ion interaction. Exchange correlation effects are treated by local density approximation \cite{lda} and freedom of spin is not considered.

\section{Results and discussion}
To determine the atomic configuration of the interface, I first perform structural optimization using a thin film model in which the dangling bonds of silicon atoms in the bottommost and topmost layers are terminated by hydrogen atoms \cite{comment1}. The numbers of atoms in the models are listed in Table \ref{tbl:atoms_and_energy}. The periodic boundary condition is imposed in the $x$ and $y$ directions, while the isolated boundary condition is employed in the $z$ direction. The vacuum region above and beneath the thin film is $\sim$ 6.5 \AA. I relax all the atoms in model (a) to determine the thickness of the thin film and then execute structural optimizations for the other models. For models (b), (c), (d), (e) and (f), the Si atoms in the first bottommost, second bottommost, and topmost layers, and the H atoms passivating the dangling bonds of the Si atoms in both the surface layers are frozen at the positions obtained by model (a) during structural optimization. Using the above parameters, formation energies of the interface defects are calculated for their optimized geometries. The fifth column of Table~\ref{tbl:atoms_and_energy} shows the formation energies $E_f$ defined as
\begin{equation}
E_f=E_{(x)}+(33-N_O) \mu_O+ (N_H-12) \mu_H- E_{(a)}.
\end{equation}
Here, $E_{(a)}$, $E_{(x)}$, $\mu_O$, and $\mu_H$ are the total energies of model (a) and the other models, and the chemical potentials of oxygen and hydrogen atoms, respectively. In addition, $N_O$ ($N_H$) is the number of oxygen (hydrogen) atoms. The chemical potentials of oxygen and hydrogen atoms are set to be those of their dimers, however the definition of them does not affect my conclusion as long as I compare the energies of the models with the same numbers of $N_O$ and $N_H$. The atomic configuration of model (d) is energetically metastable and the total energy of model (d) is higher than those of models (b) and (c) by $\sim$ 0.6 eV/supercell. In addition, one can find that the hydrogen atoms can be released easily as expected from experiments by comparing the formation energy between the models with and without the dangling bond, although the quantitative evaluation is difficult because the bonding energy is affected by the definition of the chemical potential.

\begin{figure}[htb]
\begin{center}
\includegraphics{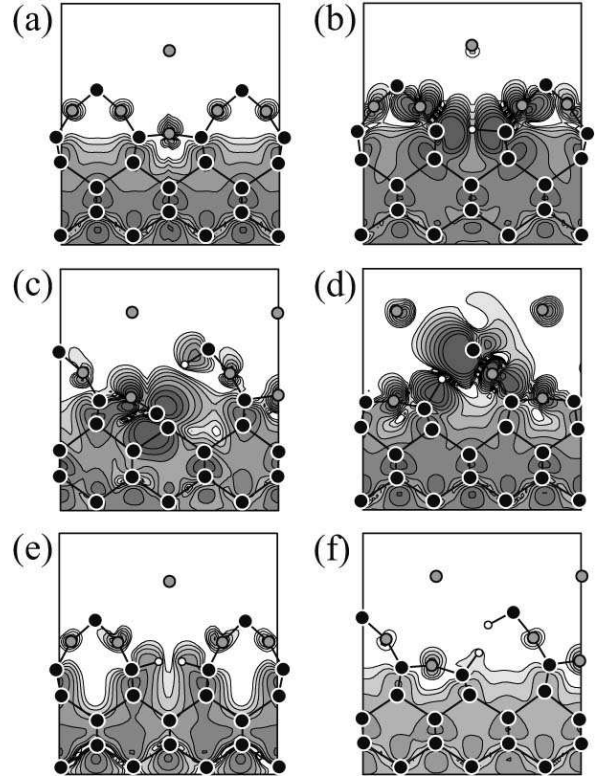}
\end{center}
\caption{Contour plots of charge-density distributions for electrons with energies between $E_F-0.5$ eV and $E_F$+0.5 eV. The planes shown are along the cross section in the (110) plane of the silicon substrate including dangling bonds. Each contour represents twice or half the density of adjacent contour lines and the lowest contour is $4.22\times 10^{-4}$ $e$/\AA$^3$. The meanings of the symbols are the same as those in Fig.~\ref{fig:fig2}. (a), (b), (c), (d), (e), and (f) correspond to models (a), (b), (c), (d), (e), and (f) in Fig.~\ref{fig:fig2}, respectively.}
\label{fig:fig3}
\end{figure}

Next, to examine the electron-conduction properties, I remove the H atoms terminating the dangling bonds of both the surface and suspend the Si/SiO$_2$ interface between aluminum jellium electrodes, as shown in Fig.~\ref{fig:fig1}. The distance between the Si atoms and the edge of the jellium electrodes is chosen to be 2.42 \AA \hspace{2mm} so as to imitate the Si(001) and Al(111) interface. Figure~\ref{fig:fig3} shows the charge-density distribution of the states between $E_F-$0.5 eV and $E_F$+0.5 eV. There are no gap states in models (a), (e), and (f), while the states due to dangling bonds are observed in models (b), (c), and (d). The charge-density distributions of the gap states are spatially localized around the defects and agree with that of the highest occupied orbital of the neutral hydrogen bridge [models (b) and (d)] \cite{hydrogenbridge} and with that of the highest occupied orbital of the $P_b$ center [model (c)] \cite{pb}. In model (c), the surface silicon atom having a dangling bond moves down to the silicon substrate so that its bond angles become planar and the $\pi$ electron forms a $P_b$ center in the silicon substrate. On the other hand, the dangling bond in model (d) forms vertically and spreads into the dioxide. The electron-conduction properties of the interface system are computed using the scattering wave functions continuing from one electrode to the other, which are evaluated by the overbridging boundary-matching method \cite{icp,obm} under the semi-infinite boundary condition in the $z$ direction. The leakage current of the system at the limits of zero temperature is evaluated using the Landauer-B\"uttiker formula \cite{buttiker}
\begin{equation}
\label{eqn:current}
I=-\frac{2e}{h}\int^{E_2}_{E_1} T(E) dE,
\end{equation}
where $T(E)$ is the transmission probability of the incident electron from the lower electrode with the energy $E$. The leakage current is computed by the integration of the tunnel current over the range from $E_1=E_F-0.5$ eV to $ E_2=E_F+0.5$ eV, where $E_F$ is the Fermi energy \cite{comment2}.

The sixth column of Table~\ref{tbl:atoms_and_energy} shows the leakage current ratio in which the current of the perfect interface is set to be unity \cite{comment3}. Note that the passivation of the dangling bonds arising from the oxygen vacancy by the hydrogen molecule markedly reduces the leakage current. On the other hand, the current increases with the presence of dangling bonds at the interface. In particular, the current in model (d) is 534.3 times larger than that in model (a) and the defect in model (d) is expected to cause serious problems in the practical devices.

\begin{figure*}[htb]
\begin{center}
\includegraphics{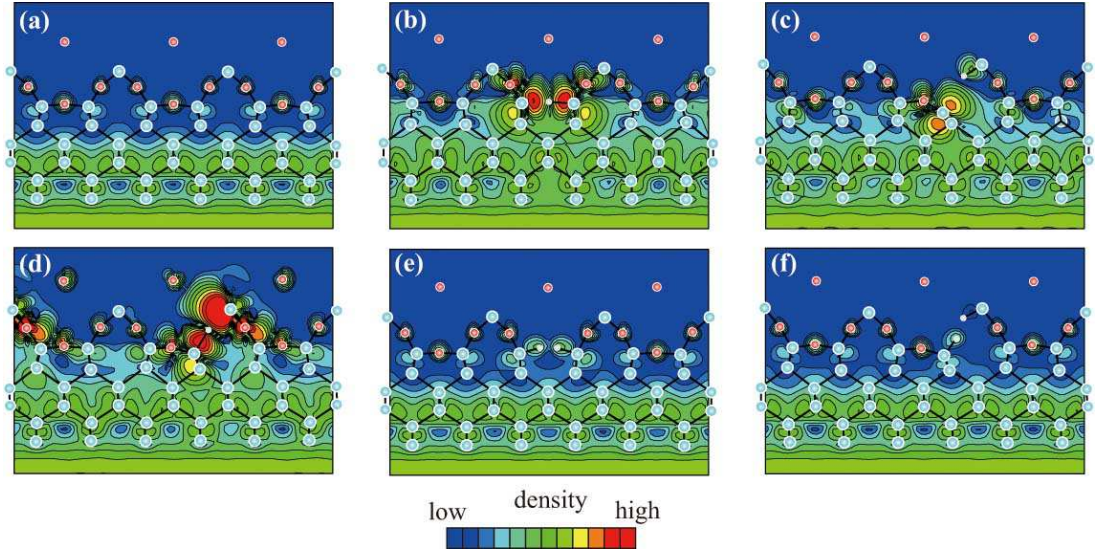}
\end{center}
\caption{(color) Charge-density distributions of scattering wave functions with the Fermi energy. Each contour represents twice or half the density of adjacent contour lines and the lowest contour is 4.72 $\times 10^{-5}$ {\it e}/eV/\AA$^3$. (a), (b), (c), (d), (e), and (f) correspond to models (a), (b), (c), (d), (e), and (f) in Fig.~\ref{fig:fig2}, respectively. Light blue, red, and white balls represent silicon, oxygen, and hydrogen atoms, respectively.}
\label{fig:fig4}
\end{figure*}

\begin{figure}[htb]
\begin{center}
\includegraphics{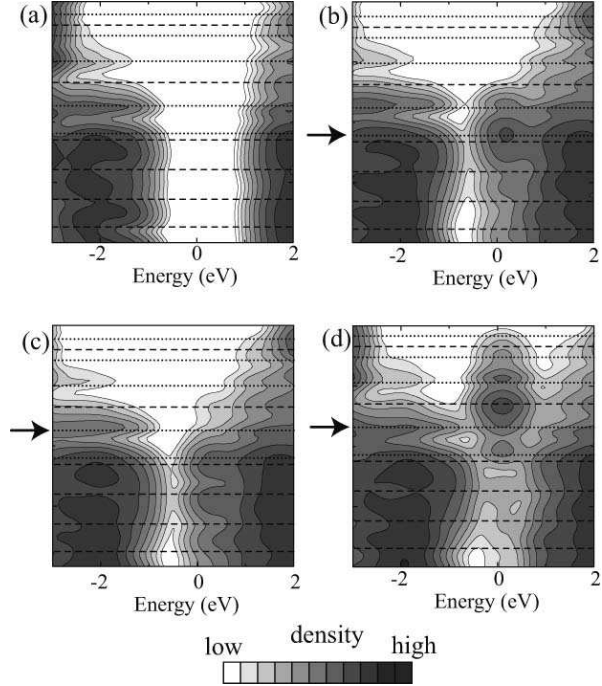}
\end{center}
\caption{Distributions of DOS integrated on plane parallel to interface as functions of relative energy from the Fermi energy. Zero energy is chosen to be the Fermi energy. Each contour represents twice or half the density of adjacent contour lines and the lowest contour is 1.39 $\times 10^{-5}$ {\it e}/eV/\AA. (a), (b), (c), and (d) correspond to models (a), (b), (c), and (d) in Fig.~\ref{fig:fig2}, respectively. Dashed and dotted lines represent the vertical positions of silicon and oxygen atomic layers, respectively. Arrows denote the oxygen atomic layer from which an oxygen atom is extracted.}
\label{fig:fig5}
\end{figure}

To explore the relationship between the presence of dangling bonds and leakage current in detail, I plot in Fig.~\ref{fig:fig4} the charge-density distribution of the scattering wave functions with the Fermi energy. Most incident electrons from the Si substrate are reflected at the Si/SiO$_2$ interface. In addition, the characteristics of the charge-density distribution around the defects completely agree with those of the gap state in Figs.~\ref{fig:fig3}(b), \ref{fig:fig3}(c), and \ref{fig:fig3}(d). Thus, the incident electrons penetrate through the gap states. On the other hand, I could not find the evidence that the dangling bonds terminated by the hydrogen atoms cause the leakage although the charge-density distributions of the scattering wave functions look similar with those in Figs.~\ref{fig:fig3}(e) and \ref{fig:fig3}(f). The dangling bond states energetically shift up to the valence band or down to the conduction band by the passivation so that the states do not contribute to the leakage. Figure~\ref{fig:fig5} shows the densities of states (DOS) of models (a), (b), (c) and (d), which are plotted by integrating them on the plane parallel to the interface following $\rho(z,E)=\int |\psi(\mbox{\boldmath$r$},E)|^2 d\mbox{\boldmath$r$}_{||}$. The width of the energy band gap at the Fermi level around the interface region in model (a) agrees with that obtained by the other first-principles calculation, in which more thick Si and SiO$_2$ films are directly bonded \cite{kaneta}. This implies that thickness of both the film in my model is sufficient to eliminate unfavorable effect from the electrode. Most electrons of the gap states in model (b) accumulate beneath the interface atomic layer because the direction of the dangling bond is parallel to the interface. In Fig.~\ref{fig:fig5}(c), the two peaks in the distribution of the DOS at the Fermi level is the evidence that the gap state of model (c) comes from the $P_b$ center, which is composed of the $\pi$ electrons. However, the state locates spatially deep inside the silicon substrate and does not contribute to the leakage significantly. In contrast, the charge-density distribution of the gap state in model (d) spreads along the direction perpendicular to the interface more widely than those of the other models and penetrates into the oxide region because the hydrogen-atom bridge aligns perpendicular to the interface and the silicon atom having dangling bond exists inside the oxide. According to the problem of the tunnel effect of a potential barrier in quantum mechanics, the transmission probability exponentially decreases with an increase in the potential-barrier length. The dangling bond in model (d), which locates in the energy band gap of the SiO$_2$ film and spreads in the direction perpendicular to the interface, results in a significant increase in leakage current. Thus, it degrades the performance of the oxide film as an insulator film of metal oxide semiconductor field-effect transistors.

\section{Conclusion}
I have implemented first-principles simulations to examine the relationship between the defects at the Si/SiO$_2$ interface and leakage current. I found that the $P_b$ center and the hydrogen-atom bridge aligned parallel to the interface do not affect the leakage current significantly. On the other hand, because of the spread of the charge-density distribution of the gap state, the hydrogen-atom bridge aligned perpendicular to the interface results in the largest leakage current, and therefore, it causes dielectric breakdown. In addition, the hydrogen termination of the dangling bonds significantly reduces decrease the leakage current.

\section*{Acknowledgements}
The author would like to thank Professor Kikuji Hirose and Professor Heiji Watanabe of Osaka University and Professor Kenji Shiraishi of University of Tsukuba for reading the entire text in its original form and fruitful discussion. This research was partially supported by a Grant-in-Aid for the 21st Century COE ``Center for Atomistic Fabrication Technology'', by a Grant-in-Aid for Scientific Research in Priority Areas ``Development of New Quantum Simulators and Quantum Design'' (Grant No. 17064012), and also by a Grant-in-Aid for Young Scientists (B) (Grant No. 20710078) from the Ministry of Education, Culture, Sports, Science and Technology. The numerical calculation was carried out using the computer facilities of the Institute for Solid State Physics at the University of Tokyo, the Research Center for Computational Science at the National Institute of Natural Science, and the Information Synergy Center at Tohoku University.

\begin{table}[thbp]
\begin{center}
\caption{Number of atoms in supercell, formation energies and leakage current ratio. The models correspond to those shown in Fig.~\ref{fig:fig2}.}
\label{tbl:atoms_and_energy}
\begin{tabular}{c|ccc|c|c} \hline\hline
Model & Si & O & H & $E_f$ (eV) & Current ratio \\ \hline
(a)    & 38 & 33 & 12 & --- & 1\\
(b)    & 38 & 32 & 13 & 3.66 & 3.9\\
(c)    & 38 & 32 & 13 & 3.59 & 12.6\\
(d)    & 38 & 32 & 13 & 4.23 & 534.3\\
(e)    & 38 & 32 & 14 & 3.21 & 0.9\\
(f)    & 38 & 32 & 14 & 3.01 & 1.6\\ \hline\hline
\end{tabular}
\\
\end{center}
\end{table}


\begin{references}
\bibitem{comment1}
Jellium electrodes are not attached during structural optimization.
\bibitem{comment2}
Although the behavior of $T(E)$ with respect to $E$ is not spiky, I computed the current by the integration so as to minimize the numerical error arising from the offset of the Fermi energy, which is derived by the periodic system in the $z$ direction. However, my conclusion is not affected when the integration range changes.
\bibitem{comment3}
Since the local density approximation seriously underestimates the gap between the valence and conduction bands, which results in the overestimate of the leakage current, I discuss only the ratio of the leakage current qualitatively.
\end{references}
\end{document}